# Enhanced Breast Cancer Tumor Classification using MobileNetV2: A Detailed Exploration on Image Intensity, Error Mitigation, and Streamlit-driven Real-time Deployment

Aaditya Surya
*Viterbi School of Engineering*
*University of Southern California*
Los Angeles, United States
asurya@usc.edu

Aditya Keshary Shah
*College of Engineering*
*Rochester Institute of Technology*
Rochester, United States
adityakeshary@gmail.com

Jarnell Kabore
*College of Engineering*
*Northeastern University*
Boston, United States
jarnellkk@gmail.com

Subash Tarun Sasikumar
*School of Arts & Sciences*
*State University of New York Plattsburgh*
Plattsburgh, United States
subashtarun25@gmail.com

*Abstract—* **This research introduces a sophisticated transfer learning model based on Google's MobileNetV2 for breast cancer tumor classification into normal, benign, and malignant categories, utilizing a dataset of 1576 ultrasound images (265 normal, 891 benign, 420 malignant). The model achieves an accuracy of 0.82, precision of 0.83, recall of 0.81, ROC-AUC of 0.94, PR-AUC of 0.88, and MCC of 0.74. It examines image intensity distributions and misclassification errors, offering improvements for future applications. Addressing dataset imbalances, the study ensures a generalizable model. This work, using a dataset from Baheya Hospital, Cairo, Egypt, compiled by Walid Al-Dhabyani et al., emphasizes MobileNetV2's potential in medical imaging, aiming to improve diagnostic precision in oncology. Additionally, the paper explores Streamlit-based deployment for real-time tumor classification, demonstrating MobileNetV2's applicability in medical imaging and setting a benchmark for future research in oncology diagnostics.**

## I. INTRODUCTION

Breast cancer is a particular variant of carcinoma distinguished by its origin in the lining of breast tissue, leading to tumor formation with potential metastatic tendencies. It is typically shown to predominantly occur in women and holds the unfortunate title as the second most common adenocarcinoma within this demographic. The World Health Organization (WHO) suggests that over 99% of breast cancer diagnoses concern women, highlighting the severe gender-centric nature of this disease [1].

Amidst this regrettable backdrop, it is noted that it is of vital necessity to timely diagnose such cases of cancer promptly, as the survival rates of patients are largely contingent on how expeditiously a diagnosis can be found. Hence, the paramount purpose of this study is to create a model that robustly and efficiently produces correct classifications for a given ultrasound scan.

The integration of deep learning into breast cancer research has shown promise that AI models can offer enhanced patient examination procedures and classify diagnoses with both speed and accuracy [2][3]. Our study leverages a dataset of breast ultrasound images, meticulously curated by Walid Al-Dhabyani, Mohammed Gomaa, Hussien Khaled, and Aly Fahmy. This invaluable resource proves to be the foundation upon which the following machine learning models will be trained, tested, and evaluated [2].

The particular dataset utilized for this research has classified types of breast cancer tumors into three categories: normal, benign, and malignant. A "normal" classification indicates the absence of any tumor detected in the image, hence indicating healthy breast tissue. A "benign" classification, on the other hand, does indicate that a growth is present but shows that the growth is non-cancerous and does not possess the ability to spread or invade surrounding tissue [2]. The last classification the data presents is malignant tumors, which prove to be the most concerning of the three as it indicates cancerous growths that have the potential to invade nearby tissue and metastasize to distant organs [2].

Table.1. Classification and Description of Tumor Types

| Class | Label | Description |
|---|---|---|
| 0 | Normal | Corresponds to healthy tissue without any abnormalities or tumor growth. |
| 1 | Benign | Refers to a non-cancerous tumor that does not invade nearby tissues nor spread to other parts of the body. |
| 2 | Malignant | Represents cancerous tumors that can invade nearby tissues and spread to other parts of the body. |

The data is utilized through Google's MobileNetV2 architecture for classifying breast cancer tumors into normal, benign, and malignant categories as shown in Table 1. An important acknowledgment to make are the limitations: hardware and software complexity. The hardware available is not at the forefront of technology and is rather commercial, ultimately restricting the model's full capabilities. Additionally, the model complexity was affected through the same issue as the memory allocation problem cannot be overlooked. A lower degree of memory leads to smaller batch sizes and directly affects the models ability to recognize more abstract elements in the data. The last limitation was the low quantity of data that was available to train the model.

## II. AIMED CONTRIBUTIONS AND RESEARCH QUESTION

The paper wants to answer the question, or rather the problem, of how the general public is supposed to access machine learning models. Up until the recent exponential increase in popularity of artificial intelligence and machine learning due to the efforts of OpenAI, many models were left in the hands of giant corporations who were the only ones looking into profit off of them. Even now, it is still difficult to find a trusted model for one to use to check for health purposes. The paper is looking into this issue and trying to offer up an idea of what could be done with more time and proper funding.

The full extent of the paper reaches an online deployment of the model bridging the gap between the academic bubble and into the general public. The implications of the model lies closer to the health care sector rather than the machine learning sector as it has a wider impact on cancer research and solutions. Due to this, the goal is to make a very user friendly model for people to be able to use as a resource if need be. It is hoped that this will lower the amount of false diagnosis' across the nation and get people the help they need early on rather than later.

## III. LITERATURE REVIEW

### A. Foundational Techniques and Approaches in Breast Cancer Diagnosis

[3] GF. Stark et al. embarked on a comprehensive journey, analyzing an array of machine learning models to enhance the accuracy of breast cancer diagnosis. Their work positioned the traditional Gail model, a cornerstone in breast cancer risk assessment, against modern machine learning frameworks. This comparative evaluation is especially significant as it provides an empirical benchmark for model performance, emphasizing the evolution and transformative potential of AI in medical diagnostics. Researchers can draw from Stark's methodology to juxtapose traditional and modern diagnostic methods, ensuring that advancements in AI-driven systems are grounded in established medical knowledge.

[5] Ahmad et al. focused on the intersection of data mining techniques and breast cancer prediction. Their research emphasized the profound capabilities of data mining, offering a systematic methodology to extract relevant patterns and insights from vast data sets. By merging these techniques with machine learning models, they paved the way for a more sophisticated understanding of breast cancer diagnostics. For researchers aiming to leverage breast cancer datasets, especially intricate ones like ultrasound datasets, integrating such data extraction techniques can lead to more enriched and informed model training processes.

The intersection of medical knowledge and machine learning models is the core of the paper, which answers the question as to why papers about traditional medical techniques and modern machine learning techniques were cited. Transfer learning techniques were a great help in achieving this as it allowed for the use of a pre-trained model of a similar task (identifying tumors in breast cancer) as the starting point for a new model. This made the model already have an idea of what to do and how to do it.

### B. Advanced Classification and Transfer Learning Techniques

[6] Gulzar, Yonis highlights in their paper an implementation of transfer learning using MobileNetV2 for fruit image classification. Though this may not be a direct implementation of transfer learning within the context of the medical field, it promptly addresses and describes how convolutional neural networks can be enhanced by pre-trained models to accurately classify images. The paper utilized a dataset of 26,149 images across 40 classes of fruits and introduced a modified MobileNetV2 incorporating a customized head to increase model robustness. The modified model leveraged transfer learning from MobileNetV2 and achieved an accuracy of 99%. When compared with other architectures the transfer learning model consistently shows to outperform in image classification.

Another instance of transfer learning is through the work of [7] Praba Hridayami. They effectively showcased the merits of utilizing the pre-trained VGG16 model within their Convolutional Neural Network (CNN) framework. Their approach, resulting in a high accuracy rate, underpins the importance of leveraging transfer learning techniques in medical imaging classification. Given the intricate patterns and nuances present in breast ultrasound imaging, integrating deep, pre-trained models can offer a solid foundation, optimizing the model's ability to discern between benign, malignant, and normal patterns. Hridayami's work underscores the need to balance depth and adaptability in machine learning architectures for high-stakes tasks such as cancer detection.

A deeper implementation of a cancer prediction model can be seen through [8] Weiming, Mi et al. in their multi-class classification study of breast digital pathology also utilized deep learning for cancer prediction. This study in particular stands out as it diverges from a simple binary classification model (such as normal vs. tumor or benign vs. malignant) and instead introduces a more profound multi-class classification system. Through employing a dual-stage architecture the study classifies breast digital pathology images into four categories: normal tissue, benign lesion, ductal carcinoma in situ and invasive carcinoma. This research proves to be a fundamental basis to building a multi-class classification system that is pertinent to the study at hand.

Thus, the machine learning model is able to utilize transfer learning techniques, but in order to make sure that the model does not go astray evaluation metrics were relied upon. The evaluation metrics ensure the models accuracy and repeatedly checks whether or not the model is accurately identifying benign, normal, or malignant tumors.

### C. Evaluation Metrics and their Importance

[9] Vakili Meysam, et al. in their paper provide an analysis of different evaluation metrics for classification algorithms. Their paper serves as a meta-analysis, examining a plethora of different metrics that encompass precision, recall, f1-score, accuracy, confusion matrices, and ROC-AUC scores. In the notably instructive section 3.3 of the paper, the authors state the definitions and formulas for each metric and their respective benefits and detriments in terms of evaluating classification models. For the purpose of this study, this paper's exhaustive overview is invaluable as the choice of an evaluation metric for the model presented below can heavily influence the interpretation of the model's efficacy.

In addition to ROC-AUD scores, another metric of accuracy was researched. This creates a more holistic evaluation score as multiple metrics will help discern unique limitations within the model. [10] Chicco, Davide, and Giuseppe Jurman in their paper describing the comparison of the Matthews Correlation Coefficient (MCC) with other relevant metrics shows the advantages MCC has in evaluating classification models. The paper compares MCC critically with F1 score and Accuracy which are often the most commonly used metrics for CNN models. While accuracy and F1 score have been popular over the past few decades, the paper shows how often they present overoptimistic results, especially in the face of imbalanced datasets. The study describes that MCC, on the other hand, provides a significantly more robust evaluation by accounting for true positives, false negatives, true negatives, and false positives while adjusting proportionally for positive and negative elements in the dataset. Given the study, leveraging MCC as an evaluation metric offers a rigorous and holistic understanding in classification studies, ensuring the authenticity of diagnostic outcomes.

*D. Optimization and Hyperparameter Tuning in Neural Networks*

A hyperparameter is an external configuration variable that are used to manage the training of machine learning models. Examples of this include the number of nodes/layers in a neural network and the number of branches in a decision tree. They determining the significant features such as the model architecture, learning rate, and model complexity.

Due to the importance of hyperparameters, it is imperative to understand the current breakthroughs in optimizing and manipulating them for maximum effect. [11] Saleh, Hager, et al. in their research in different deep learning approaches towards breast cancer diagnosis utilize Keras-Tuner in optimizing their architecture. The paper offers an advanced methodology using an optimized Recurrent Neural Network (RNN). The Keras-Tuner in the paper is described to be chosen for its flexibility, user-friendliness, and its streamlined manner in optimizing hyperparameters without the cumbersome trial and error process. In this study, the tuner optimized different dropout rates between 0.1 to 0.9 and showed a significant improvement in diagnostic accuracy with the optimizations.

IV. METHODOLOGY

*A. Data Collection*

The data source is a series of photos and data of a baseline breast ultrasound taken from women between the ages of 25 and 75 collected in 2018. The dataset comprises 1576 images in PNG format, classified into three distinct categories: normal, malignant, and benign. A normal scan indicates the absence of any abnormal growths in the breast. A scan classified as malignant indicates the presence of harmful growths, while a benign classification signifies the detection of harmless growths.

To further understand the distribution of the dataset, Figure 1 presents a pie chart detailing the numerical and percentage distribution of each class. Specifically, there are 265 normal instances, 891 benign instances, and 421 malignant instances. This visualization is pivotal for our paper as it underlines the inherent class distribution and aids in understanding potential challenges related to class imbalances. Addressing and understanding these imbalances is crucial to ensure the reliability and generalizability of deep learning models trained on this data.

The data was partitioned into three subsets: training data (consisting of 499 images), validation data (125 images), and test data (156 images). This structured division aids in systematically training, tuning, and evaluating the deep learning models. The data was meticulously collected by Walid Al-Dhabyani et al. [2] using the LOGIQ E9 ultrasound and LOGIQ E9 Agile ultrasound system at the Baheya Hospital for Early Detection & Treatment of Women's Cancer, Cairo, Egypt. The primary motivation behind this collection was to offer a rich dataset for individuals and researchers keen on delving into the realms of deep learning applications in medical imaging.

**Figure. 1. Distribution of Classes in the Dataset**

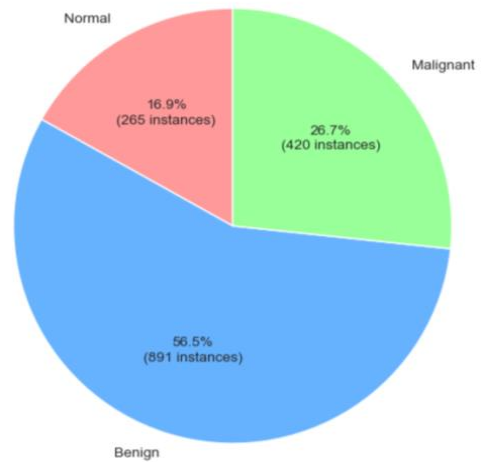

*B. Data Preprocessing*

Before the data was introduced to our model, an extensive preprocessing phase was undertaken to ensure its cleanliness and applicability for the research's objectives.

1. Rescaling: Pixel values of each image were rescaled to a range between zero and one, ensuring uniformity across the dataset.
2. Rotation: Images were randomly rotated by up to 20 degrees in any direction. This step prevents the model from becoming overly sensitive to the orientation of input scans, accommodating potential variations in future datasets.
3. Positional Augmentation: The heights and widths of the images were randomly adjusted. This process captures those cases where the growth could be off-center, ensuring the model is robust to all positional variations.
4. Shear Transformation: The images underwent shearing to recognize objects even if slightly distorted. This step mitigates issues where scans may be distorted due to operational errors, be it human-induced or machine-related.
5. Flipping: Images were subjected to horizontal flips, enhancing the model's ability to discern objects regardless of their horizontal orientation.
6. Fill Function: To ensure the integrity of images post-transformation, a fill function filled in any missing

pixels that might have resulted from the rotation or flipping processes.

These augmentations weren't applied statically. Instead, they were applied dynamically and randomized during each epoch. Rather than creating a fixed set of augmented images at the outset, our approach involved the random application of these augmentations every time an image was processed during training.

After processing the images, the model underwent training for 50 epochs with batches of 32 images. Each epoch exposed the model to 50 uniquely augmented images, cumulatively resulting in 2,500 distinct images over the training phase. To elucidate, the batch size indicates the number of training instances used in one iteration. For our training, sets of 32 images were randomly picked, processed, and then used to adjust the model's weights through the backpropagation algorithm. The choice of 32 as the batch size struck a balance between computational efficiency and data clarity. Additionally, the induced variability in the weight updates, inherent with smaller batches, acted as a regularization method, potentially assisting in avoiding local minima and promoting a more generalized model.

The entire dataset was apportioned into training, validation, and test subsets, as previously detailed. Post-processing snapshots of the data are presented in Figure 1.2. It's noteworthy that while the training data experienced the full suite of augmentations, the validation set was solely rescaled. This distinction ensures the validation set remains untouched by augmentations, offering a genuine performance metric. The test set remained isolated from this process and was reintroduced at the culmination of the research to assess the model's proficiency.

**Figure. 2. Ultrasound Breast Cancer Scans**

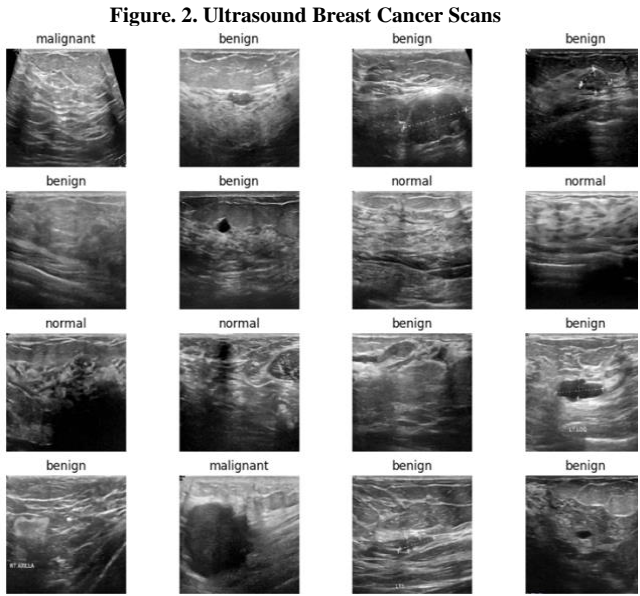

## V. MODEL ARCHITECTURE

### A. Flowchart of Model Development

The subsequent flowchart provides a succinct visual representation of the stepwise progression undertaken in our study's model development phase. Beginning from data collection, it charts a clear trajectory through preprocessing, model training, validation, and, ultimately, testing. This graphical representation facilitates a quick understanding of the workflow and ensures clarity in the process of our deep learning implementation for breast cancer classification.

**Figure. 3. Flowchart of Model Development**

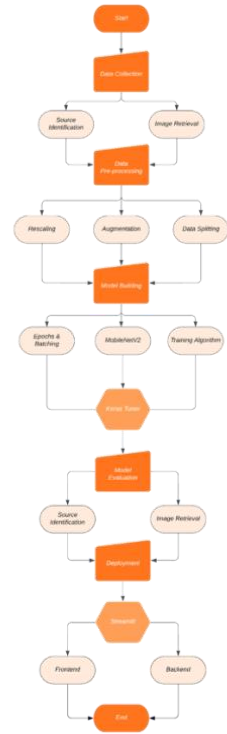

### B. Transfer Learning with Mobile Net V2

Transfer learning has emerged as a transformative approach to creating convolutional neural network classification models. By leveraging a pre-trained model, we can expedite the learning process and subsequently refine it to the specificity of our dataset. The architecture selected for this endeavor was the "MobileNetV2".

Developed by Google researchers, MobileNetV2 is renowned for its efficiency, making it apt for mobile and edge devices. Trained on the expansive 'ImageNet' dataset, MobileNetV2's convolutional base is equipped with a nuanced understanding of a myriad of image categories. This pre-training offers an invaluable starting point. However, given the specificities of our dataset vis-à-vis ImageNet's multi-class nature, certain adaptations were deemed necessary.

To bridge the gap between ImageNet classes and our data categories, the top layers of the MobileNetV2 model, which are more task-specific, were excluded to ensure they didn't impose any biases from the pre-trained weights [6]. The architecture was then fine-tuned to accommodate our dataset with an input shape of (150, 150, 3), aligning it with the unique morphology of our ultrasound image scans.

After the model's tailoring, an important step was to freeze its initial layers. Convolutional layers in architectures like MobileNetV2 present a hierarchical stratification of visual features. Initial layers predominantly discern basic image constructs like edges and textures. As one progresses deeper into the network, layers discern more intricate, dataset-specific features. By deploying transfer learning, it's prudent to freeze the initial layers. This retains the generalizable features learned from the ImageNet dataset, ensuring that they don't unduly influence the model's fine-tuning on our specialized dataset. This strategy ensures a harmonious blend of general visual understanding and specific feature recognition, pivotal for our research's goals.

## VI. Hyperparameter Tuning and Calibrating Deeper Layers with MobileNetV2

Fine-tuning models, especially when leveraging the transfer learning approach, ensures that the trained architecture adapts astutely to the unique nuances of the project-specific dataset rather than merely inheriting patterns from a generic source like ImageNet. This calibration is vital for precise interpretation, especially when transitioning datasets.

Regarding MobileNetV2's architecture, specific nuances were attended to during this research. Batch Normalization layers, though often inherent in many deep learning models, were purposely excluded from this iteration. This is done to avoid distribution statistic discrepancies. In essence, pre-trained models bring along with them specific statistics - the mean and variance from their originating dataset. Using these exact metrics could muddle training when applied to a divergent dataset.

Certain layers intrinsic to MobileNetV2 underwent changes as well. While MobileNetV2 is designed for efficiency with depth wise separable convolutions, our task required some tweaks for optimal results. By bypassing some of its traditional configurations, the model was augmented with a flattened layer, making the output from the previous layers amenable to a dense structure. This flattened layer was succeeded by a dense layer containing 1024 nodes, a dropout layer to counter overfitting, and another dense layer with three nodes, aligned with our project's categorical outcomes.

The "Flatten" layer deserves emphasis. It transitions feature maps from the preceding pooling layer into a singular dimensional vector. Such an alteration is pivotal, bridging the convolutional output to the dense layers. The introduced 50% dropout layer wasn't a capricious decision either. It was interwoven to mitigate overfitting and to ensure a balanced neuron interplay, preventing any undue dependencies and thereby facilitating balanced data processing.

Complementing manual calibrations, the study made use of the Keras Tuner for a more systematic hyperparameter optimization. A popular choice for fine-tuning CNN models, the Keras Tuner discerns optimal configurations, from learning rates and dropout percentages to node quantities in dense layers and the most congruous activation functions [11]. Such a meticulous endeavor ensures the MobileNetV2 model's peak performance for the ultrasound images under scrutiny.

## VII. Evaluation metrics for medical diagnosis

The rigor of evaluating the performance of deep learning models in medical diagnostics is unparalleled. Not only does a misclassification bear direct consequences, but it also stresses the need for transparent and comprehensive metrics. To this end, an array of metrics was diligently selected to evaluate our MobileNetV2 model tailored for tumor image classification.

### A. Classification Accuracy and Precision

Accuracy represents the proportion of correct predictions to the total predictions. It's vital in medical imaging as a basic measure of a model's capability to correctly identify tumor classifications.

$$Accuracy = \frac{Number\ of\ correct\ predictions}{Total\ number\ of\ predictions}$$

Precision, particularly in medical diagnostics, where false positives can lead to unnecessary treatments or interventions, precision's value cannot be overstated. It measures the consistency of the model's outcome for a specific classification.

$$Precision = \frac{True\ Positives}{True\ Positives + False\ Positives}$$

### B. Loss Value

Especially crucial in multi-class classification tasks in medical imaging, where errors can have grave implications. The Categorical Cross-Entropy loss value function was employed to direct weight adjustments during the training phase, bridging the gap between predicted and actual values.

### C. AUC (Area Under Curve)

AUC, as related to the ROC curve, gauges model performance over various threshold values [9]. In diagnostics, where the binary distinction between malignant and benign tumors is pivotal, AUC's role is magnified. The computation considers the balance between the true positive rate (recall) and the false positive rate:

$$TPR\ (True\ Positive\ Rate) = \frac{True\ Positives}{True\ Positives + False\ Negatives}$$

$$FPR\ (False\ Positive\ Rate) = \frac{False\ Positives}{False\ Positives + True\ Negatives}$$

### D. Matthews Correlation Coefficient

An integral metric when dealing with datasets showcasing uneven class distributions, often a reality in medical data sets where certain conditions are rare [9]. MCC's formula encompasses all facets of the confusion matrix:

$$MCC = \frac{TP \times TN - FP \times FN}{\sqrt{(TP + FP) \times (TP + FN) \times (TN + FP) \times (TN + FN)}}$$

### E. Precision-Recall Area Under Curve (PRAUC)

Vital for cases where imbalances in data are the norm or when precise detection (like rare diseases) is the focus. PRAUC embodies a model's finesse in harmonizing precision and recall across assorted thresholds.

By incorporating these metrics into the evaluation phase, the study seeks to guarantee not just the model's precision but also its reliability and reproducibility, especially in the high-stakes arena of medical diagnostics. Continuous monitoring of these metrics throughout the training and validation stages facilitated the progressive refinement of the MobileNetV2 model, ensuring its readiness for pivotal medical applications.

## VIII. Results and Discussions

This section offers an in-depth comprehensive analysis of the convolutional neural network (CNN) model's overall performance within the area of breast ultrasound class. Our study remains intently targeted at achieving the vital objective of correctly discerning normal, benign, and malignant ultrasound pictures using the model. In this section, we delve into the evaluation of the model's efficacy in achieving the objective of our study. Figure 4 provides all essential model evaluation metrics interpreted in this section.

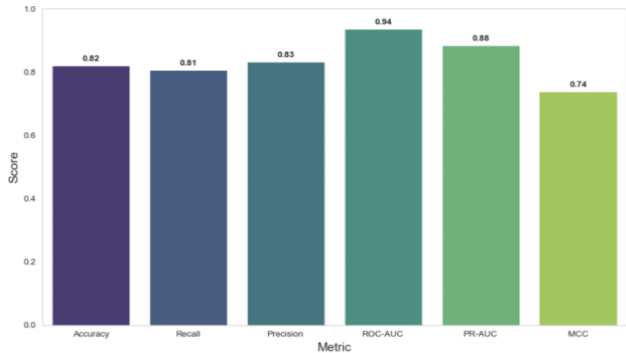

Figure. 4. Essential Model Evaluation Metrics

### A. Accuracy

The derived accuracy metric, quantified at 0.82, stands as a testimony to the model's prowess in correctly classifying approximately 82% of the analyzed ultrasound samples. This metric holds profound implications in the context of tumor classification, and the capability to reap such an accuracy is pivotal as it substantiates the model's ability as a reliable tool in the domain of medical imaging and diagnosis. This degree of precision is crucial for supporting clinical judgment and therapeutic direction, especially when it comes to spotting possible cancers. In order to ensure early diagnosis and proper patient management, the model's ability to provide accurate classification is crucial.

### B. Precision

The model's ability to correctly identify 83% of anticipated positive samples, as indicated by a precision of 0.83 shown in Fig. 1. This precision metric highlights its effectiveness in lowering the incidence of false alarms. Such precision has significant ramifications, especially in the delicate area of medical diagnosis, reducing unnecessary procedures and consequent patient emotional discomfort.

### C. Recall

The recall metric (0.81) demonstrates the model's exceptional ability to accurately identify and capture 81% of positive cases in the dataset. This proficiency in minimizing false negatives is crucial for medical diagnostics, as missing true positive cases can have significant consequences. This high recall data point enables the timely detection of potential malignancies or abnormalities in ultrasound images, improving patient safety and medical decision-making. The model's potential in medical practice, tumor classification using Deep Learning, and healthcare standards enhancement further emphasize its value.

### D. ROC-AUC

The ROC-AUC score of 0.94 indicates excellent model performance, with a curve area approximating 1. The Receiver Operating Characteristic (ROC) curve represents the model's true positive rate against its false positive rate and the model's discriminatory power. Whereas the Area Under the Curve (AUC) measures its effectiveness in distinguishing positive and negative cases. The high ROC-AUC score high underscores the model's proficiency in tumor classification, highlighting its potential clinical utility. Figure 5 OC scores across each class provides critical insights into the model's distinguishing power. Class 0 Performance: Notably, class 0 shines distinctly, registering a remarkable ROC score of 0.97. This score signifies the model's superior aptitude to discern instances of class 0 from the broader dataset. Such a high ROC score elucidates the model's competency in curtailing false positives while concurrently maximizing the true positive rate. This ability is paramount in the realm of medical diagnostics.

Figure. 5. ROC Curves for Multi-Class Classification

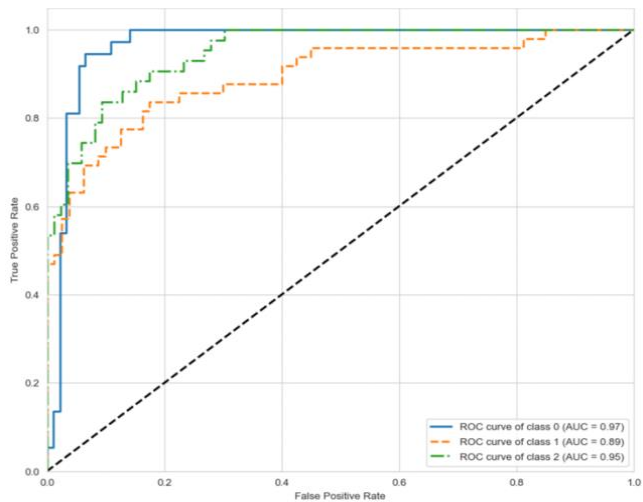

### E. PRAUC

The Precision-Recall Area Under the Curve (PR-AUC) score of 0.88 indicates the model's performance across precision and recall thresholds, providing valuable insight into imbalance datasets or clinically significant classes. It evaluates the model's ability to balance precision (identifying positive cases) and recall (capturing all positive cases). This high PR-AUC score enhances the model's suitability for real-world clinical applications and strengthens its value in Deep Learning-based Tumor Classification.

The Precision-Recall (PR) curves for each class in Figure 6 indicate AUC values of 0.86, 0.88, and 0.91 for classes 0, 1, and 2, respectively. These curves exemplify the intricate equilibrium between accurately identifying positive instances and minimizing false positives.

The PR curves offer an intuitive representation of the interplay between precision and recall across diverse thresholds. The associated AUC values capture this balance statistically, aiding in the selection of optimal decision thresholds aligned with specific clinical requirements.

The model's precision-recall dynamics are further illustrated by the PR-AUC scores for each class: 0.86 for class 0, 0.88 for class 1, and 0.91 for class 2. These scores adeptly highlight the model's refined performance in achieving the best true positive detection while judiciously limiting false positives.

Figure. 6. Precision Recall Curves for Multi-Class Classification

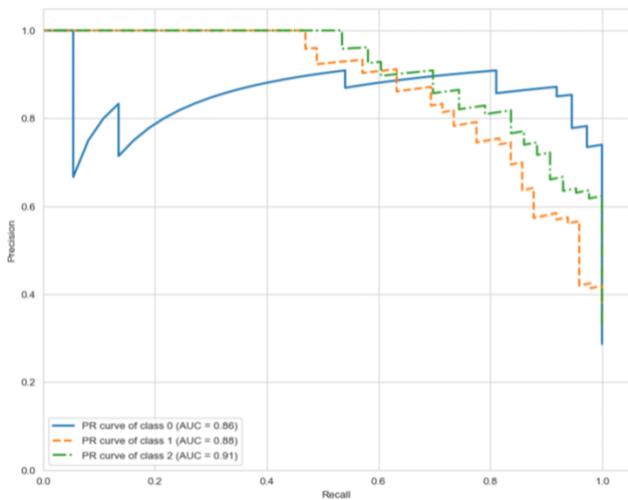

### F. MCC (Matthews Correlation Coefficient)

The Matthews Correlation Coefficient (MCC) score of 0.74 indicated the model's strong correlation between observed and predicted classifications, encompassing true positive, true negative, false positive, and false negative cases. This metric evaluates the model's performance across all four outcomes in medical diagnostics, where sensitivity and specificity are crucial. A high MCC indicates a stronger agreement between predictions and the ground truth, indicating consistent and reliable results [10]. This score demonstrates the contributional opportunity to accurately diagnose while simultaneously aligning with medical practice demands ultimately validating its role in advancing deep learning-based tumor analysis.

The reason why this score is the lowest is because the Matthew's Correlation Coefficient regards the negative class samples (true negative and false negative) highly. This highlights that the model is not always accurate when discerning when someone does not have breast cancer leading to an increase in claims of malignant tumors when the tumors are actually benign or normal.

## IX. MODEL PERFORMANCE ANALYSIS

### A. Validation Analysis

To comprehensively evaluate the performance of our tumor classification model we conducted a detailed analysis of key performance metrics across training epochs. A constant rising trend is shown in Figure 7's Validation Accuracy evolution of validation accuracy over epochs. This improvement demonstrates the model's ability to recognise complex patterns in breast ultrasound pictures and modify internal representations to produce precise predictions. We display the Validation Precision versus epochs in Figure 7 It is noteworthy that accuracy shows an initial rise before leveling off in later epochs. This finding highlights the model's capability to recognize positive examples with accuracy, but it also raises the possibility of a saturation point for precision advances. Figure 7's Validation Recall shows a steady and significant increase. The model's ability to accurately identify real positive cases is demonstrated by its high recall values, which is an important characteristic in medical applications where minimizing false negatives is crucial. Lastly the Validation Loss in relation to epochs in Figure 7 The model's increased convergence and learning are indicated by the observed consistent decline in loss values. This decrease in loss is consistent with the rising trends shown in recall, accuracy, and precision, which supports the model's overall effectiveness.

**Figure. 7 Model Performance Analysis Graphs**

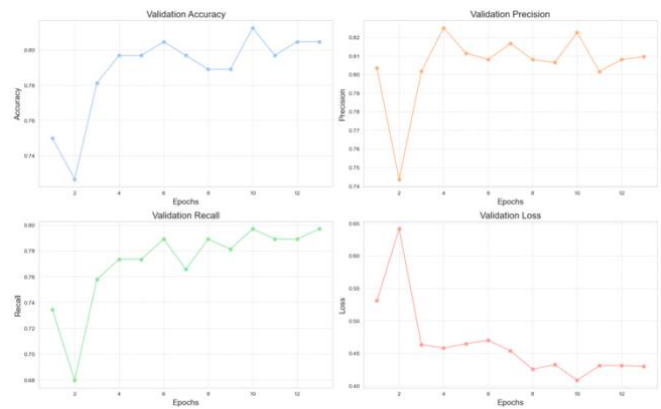

### B. Confusion Matrix

A confusion matrix is a crucial tool for evaluating the model's classification performance. It succinctly provides a thorough picture of the anticipated versus real labels, enabling a straightforward assessment of the model's advantages and weaknesses. The diagonals of the matrix stand in for accurate classifications, and certain entries show occurrences of true positives, true negatives, false positives, and false negatives.

The diagonals (35, 34, 37) for the described matrix (35, 1, 1; 6, 34, 9; 2, 4, 37) show precise classifications for classes 0, 1, and 2. The performance of the model can be improved by highlighting probable areas for improvement, such as the 9 occasions where class 1 was incorrectly categorized as class 2. The need to emphasize how crucial it is to recognize classification errors is especially clear in the field of medicine, where incorrect classification can have serious repercussions. Inaccurate classifications may result in postponed or ineffective medical measures, endangering the health and outcomes of patients.

**Figure. 8. Confusion Matrix**

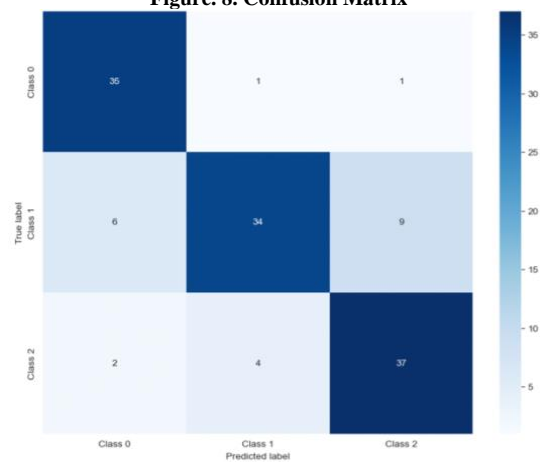

## X. IMAGE INTENSITY ANALYSIS

The distribution of image intensities, when examined across all categories, illustrates a clear right-skewed pattern. This observation suggests that most images primarily feature lower intensity values, witnessing a gradual reduction in frequency as we move towards the higher intensity spectrum. Such a distributional characteristic infers that the dataset's mean is inclined to be greater than its median, which the box plot statistics further corroborate.

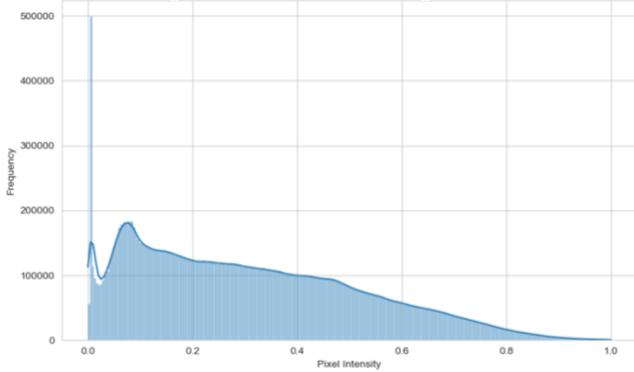

Figure. 9. Distribution of Image Intensities

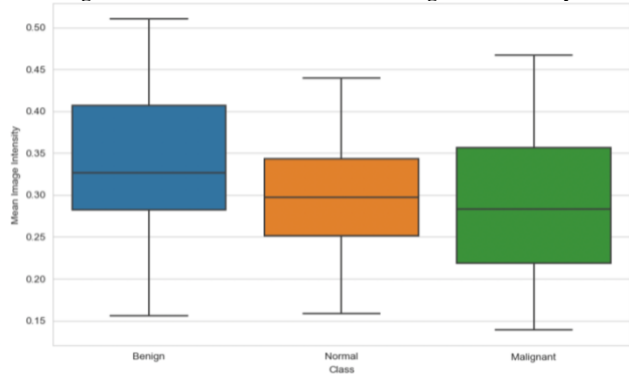

Figure. 10. Box Plot Distribution of Image Intensities by Class

Focusing on the normal image intensities, they range between 0.16 and 0.44. The central spread of data, representing 50% of these images, resides within the 0.25 to 0.34 range, emphasizing a consistent luminosity prevalent in normal images. The median intensity value is recorded at 0.30, underpinning this observation.

Benign images exhibit a broader spectrum of intensities, spanning from 0.16 to 0.51. The interquartile range, indicative of the central half of the data, stretches between 0.28 and 0.41. This expanded range, juxtaposed against the normal images, points towards an augmented variability in the illumination characteristics of benign images. Their median intensity pegged at 0.33, is slightly elevated when compared to the normal class.

Malignant images, intriguingly, present the widest span of intensities, which are contained between 0.14 and 0.47. The middle 50% of the data is ensconced between 0.22 and 0.36. The median intensity value for malignant images is charted at 0.28, indicating a tendency for these images to possess lower intensities in comparison to benign and normal images. Such trends might be emblematic of the inherent attributes or behavior of malignant tissues under imaging.

Across all categories, there is a pronounced absence of outliers, highlighting consistent imaging and processing standards in the dataset. This consistency is particularly vital in the domain of medical imaging, ensuring diagnostic dependability. While the intensity spans for individual categories exhibit overlapping tendencies, their central tendencies and dispersion metrics are distinct. Such nuances suggest that image intensity may not be the sole determinant for classifying an image, but it undeniably plays a significant role in assisting such categorization. Our analytical journey into the realm of image intensities yields crucial insights into the distinguishing features inherent to normal, benign, and malignant images. The observed disparities in the intensity profiles of these categories suggest the potential influence of either intrinsic tissue properties or imaging techniques in determining the resultant image outcomes. These findings, when harnessed effectively, could significantly elevate the precision of medical diagnostics.

## XI. ERROR ANALYSIS

In our quest to further refine and optimize our model, conducting a thorough error analysis offers an invaluable perspective. Such an analysis provides insights into where our model is faltering, allowing us to implement precise solutions

### A. Analyzing Misclassified Images

Upon reviewing the table of misclassified images, it becomes evident that certain images are consistently misinterpreted by our model. This table depicts side-by-side comparisons, showcasing the true label of an image alongside the predicted label by our model. For instance, consider an image within this table: The true label might be "Benign," but our model erroneously predicts it as "Malignant". On examining this image, we might observe that its characteristics are borderline, making it challenging even for a human expert to classify. Such a subtle differentiation point can throw off our model, causing these misclassifications. This form of visualization helps us understand the specific instances and scenarios where our model is underperforming. By taking a closer look at these misclassified images, we can infer potential reasons for these errors and strategize on how best to rectify them.

### B. Potential Reasons for Misclassification

- Data Quality: One of the fundamental pillars of any machine learning model is the quality of its training data. It's imperative to ensure that our dataset is free from label noise. Mislabeled images can cause the model to learn incorrect patterns, thereby affecting its generalization capabilities on unseen data. Regular audits and checks on the dataset.

- Data Representation: Some images might inherently possess a high degree of complexity, with subtle features playing pivotal roles in classification. If our current model architecture isn't attuned to capture these nuances, it could result in misclassifications. This calls for a reevaluation of our model's architecture and perhaps a move towards a more sophisticated design.

- Lack of Relevant Data: A balanced and diverse dataset is a cornerstone for robust model performance. If our dataset lacks ample

representation for certain classes, the model's performance for those classes would invariably suffer. While techniques like data augmentation offer some respite, there's no substitute for acquiring more real-world, varied samples for underrepresented classes.

- Complexity of the Model: While simplicity has its merits, an overly simplistic model might not do justice to a complex dataset. It might be worth experimenting by ramping up the model's complexity, incorporating more layers, or even integrating a different pre-trained model.

- Hyperparameters: Every model has a set of hyperparameters that allow it to perform optimally. Our current choice of hyperparameters like regularization rates, learning rates, or batch sizes might not be in this optimal range. Engaging in systematic hyperparameter tuning can unearth a combination that boosts model performance.

- Class Imbalance: In scenarios where certain classes overshadow others in the dataset, there's a tangible risk of the model developing a bias. This bias can skew its predictions in favor of the dominant classes. Addressing class imbalances through techniques like oversampling, under sampling, or applying class weights can pave the way for a more balanced and objective model.

While our model demonstrates commendable performance, there's always room for refinement. Through systematic error analysis and addressing the potential reasons for misclassifications, we can inch closer to a model that not only performs better but also resonates more with real-world complexities.

**Figure. 11. Misclassification Diagram**

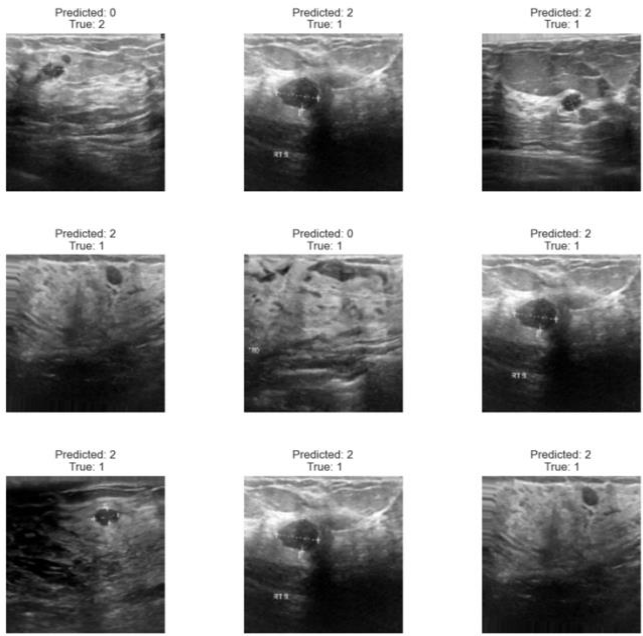

## XII. COMPARATIVE ANALYSIS OF OUR FINDINGS WITH CONTEMPORARY RESEARCH

### A. Mishra et al.'s Radiomics-based Approach vs. MobileNetV2-based Classifie

Using the same dataset to tackle breast ultrasound tumor classification, Arnab Mishra et al. applied a radiomics based approach. While similar to the research done above, the study by Mishra et al. revolved their approach to classification by heavily focusing on extracting various image features from the region of interests and incorporated a recursive feature elimination method for selecting the most pivotal features. In contrast to this approach, the MobileNetV2 model utilized for this study zeroes in on the transfer learning paradigm, specifically leveraging a pre-trained model and meticulously fine tuning it to harmonize with the dataset given.

A pivotal aspect that is pertinent to this study is the inherent class imbalance given in the dataset. To address this, Mishra et al. employed synthetic minority oversampling (SMOTE) while our study simply used random oversampling due to computational restraints. In terms of feature significance, Mishra et al.'s study revealed that the shape, texture, and histogram-oriented gradient features stood out as paramount for the classification endeavor. However, the MobileNetV2 methodology bypasses manual feature extraction by capitalizing on pretrained weights inherited from the ImageNet dataset.

Mishra et al.'s approach is radiomics-centric, emphasizing the extraction of a diverse range of image features from the region of interests and further refining them using recursive feature elimination. Their research achieved an impressive accuracy of 0.974, F1-score of 0.94, and a Matthews correlation coefficient (MCC) value of 0.959 on the BUSI dataset. In stark contrast, the MobileNetV2-based classification, which utilizes the transfer learning paradigm and fine-tunes the architecture for the specific dataset, secured an accuracy of 0.82, precision of 0.83, and recall of 0.81. Additionally, while the ROC-AUC for the MobileNetV2 method was 0.94, almost mirroring that of Mishra et al., the MCC was slightly lower at 0.74, indicating a lesser degree of reliability in the binary classifications made by the model.

Considering practical applicability, Mishra et al.'s stellar outcomes indicate a strong potential for real-world medical diagnostic applications. Simultaneously, the MobileNetV2 methodology, despite a slightly lower accuracy, suggests immense promise for real-time applications, especially on resource-constrained platforms like mobile devices, prioritizing rapid responses with substantial precision. [12]

### B. Cruz-Ramos et al.'s Hybrid Feature Fusion vs. MobileNetV2-based Classifier

Cruz-Ramos et al. ventured into the realm of breast tumor classification by employing a hybrid methodology that fused deep learning features with traditional handcrafted ones. Their Computer-Aided Diagnostic (CAD) system's core revolved around the amalgamation of DenseNet 201 architecture and traditional handcrafted features including Histogram of Oriented Gradients (HOG), ULBP, perimeter area, area, eccentricity, and circularity. The fusion process was reinforced using genetic algorithms and mutual information selection, followed by employing classifiers like XGBoost, AdaBoost, and Multilayer Perceptron (MLP).

Their study spanned two imaging modalities, mammography (MG) and ultrasound (US), leveraging datasets mini-DDSM and BUSI. Their results were indeed impressive, with an accuracy (ACC) of 97.6%, precision (PRE) of 98%, recall of 98%, F1-Score of 98%, and IBA of 95% on the aforementioned datasets. This holistic approach speaks volumes about the potential of combining traditional image processing techniques with modern deep learning architectures to obtain high precision diagnostic results.

In contrast, the research based on the MobileNetV2 architecture concentrated solely on deep learning, adopting a transfer learning paradigm and aligning the model with the specificities of the BUSI dataset. Results obtained from this method exhibited an accuracy of 0.82, precision of 0.83, recall of 0.81, ROC-AUC of 0.94, and MCC of 0.74. While both methods achieved commendable results, Cruz-Ramos et al.'s hybrid method, which combined deep and handcrafted features, exhibited superior performance metrics.

The implications of these findings suggest that while deep learning approaches like MobileNetV2 offer substantial promise, there might be untapped potential in merging traditional image processing techniques with modern neural network architectures for enhanced diagnostic precision in the domain of breast tumor classification. [13]

### C. Labcharoenwongs et al.'s Tumor Detection with YOLOv7 vs. MobileNetV2-based Classifier

Labcharoenwongs and team delved deep into breast tumor detection, classification, and volume estimation using a comprehensive deep learning approach. Their primary objective was to facilitate radiologists with automated tools that can complement their decision-making, especially when the manual analysis is influenced by radiologist skill levels and image quality. For this, they utilized the YOLOv7 (You Only Look Once version 7) architecture to detect, localize, and classify tumors from ultrasound images. Their dataset comprised 655 images, with a mix of benign and malignant samples, and they augmented this dataset with various methods such as blurring, flipping, and introducing noise.

Their results are worth noting. The YOLOv7 architecture achieved a confidence score of 0.95 for tumor detection. In terms of lesion classification, their model achieved an accuracy of 95.07%, sensitivity of 94.97%, specificity of 95.24%, a positive predictive value (PPV) of 97.42%, and a negative predictive value (NPV) of 90.91%.

On juxtaposing this with the research deploying the MobileNetV2 architecture, the following observations arise: The MobileNetV2 model exhibited an accuracy of 0.82, precision of 0.83, recall of 0.81, ROC-AUC of 0.94, and MCC of 0.74. Labcharoenwongs et al.'s model, driven by the YOLOv7 architecture, seems to achieve higher accuracy, sensitivity, and specificity.

This suggests that while the MobileNetV2 architecture is adept at handling breast ultrasound image data, architectures like YOLOv7, tailored more towards real-time object detection, might offer an edge in terms of tumor detection and classification in this specific context. The potential integration of such systems into conventional ultrasound machines, as proposed by Labcharoenwongs and team, underscores the evolving landscape of medical diagnostics and the role of deep learning therein. [14]

## XIII. Deployment

The deployment phase marks a pivotal transition in a machine learning project: from the confines of research and development to tangible, real-world applications. One of the chosen platforms for such transitions, especially in the realm of data-driven insights, is Streamlit. Recognized within the data science ecosystem, Streamlit is specialized in catalyzing the creation of interactive web applications. Its distinct advantages, such as enhanced interactivity and swift deployment capabilities, position it as a preferred choice for introducing advanced algorithms to diverse user groups. The adoption of Streamlit for this project was informed by its robust scalability which can accommodate the intricacies of complex models, combined with an interface that does not require exhaustive web development expertise.

### A. UI / UX

At the user interaction frontier, the Streamlit interface has been designed to maximize both functionality and user-friendliness. The interface facilitates the direct upload of breast cancer images, supporting a gamut of formats including PNG, JPG, and others. Upon the initiation of an upload, the application provides immediate feedback, thereby negating ambiguities. Moreover, real-time visualizations of the uploaded images enhance user interaction, with features like zooming for detailed insights. Predictions, central to the application, are rendered transparently, frequently in percentage form. These values, indicative of malignancy likelihood, are reinforced with color-coded distinctions for immediate comprehension.

### B. Backend Operations

While the front-end addresses user interactions, substantial computational operations underlie the application. To prioritize user privacy, the system is designed to process images without long-term retention. The preprocessing mechanisms are streamlined to echo the research phase methodologies, ensuring a high degree of accuracy. Response times for predictions have been optimized to be near-instantaneous, facilitating user engagement.

### C. Code Implementation

The deployment code commences with necessary imports, notably Streamlit, TensorFlow, and PIL, ensuring the required libraries are available. Once initialized, a pre-trained model, optimized for tumor classification, is loaded into the application. A function, predict_tumor(), is defined to resize, preprocess, and feed the image to the model, returning the predicted class of the tumor. Enhancing the user interface, a centralized header introduces the application, followed by prompts guiding the user to upload an image. Upon image upload, the application displays the chosen image, processes it, and presents the prediction. This deployment, while providing an immediate application of machine learning research, also lays the foundation for further refinement and expansion. Feedback loops, system integrations, and richer datasets stand as promising avenues for future iterations. Provided below is an image of our simple preliminary deployment of the created model.

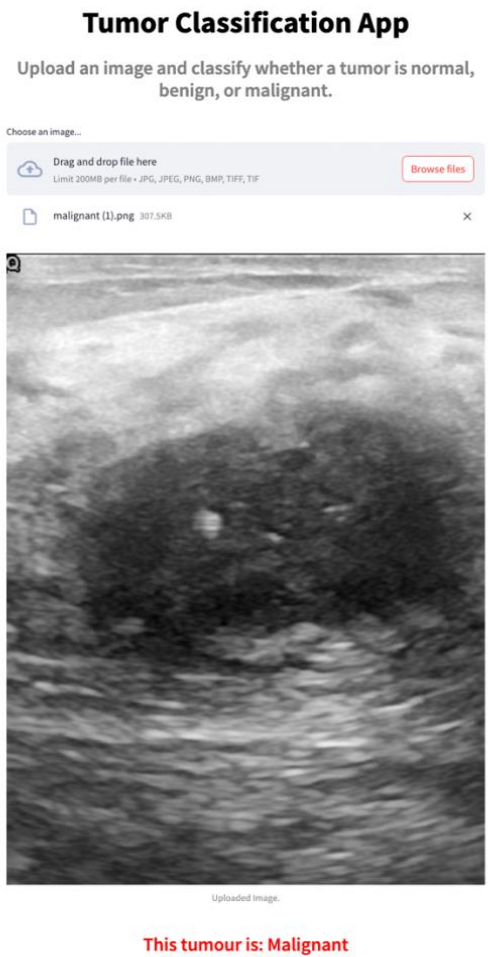

**Figure. 12. Beta Deployment UI**

## XIV. LIMITATIONS

Model development is heavily impacted by the limitations imposed by hardware and software complexities. The implementation of larger and more complex models can be restricted by a lack of computational resources, which could impede the investigation of cutting-edge architectures. Additionally, the use of intensive computational methods without necessary computational resources would limit the depth and complexity that the model's design could support. The time commitment needed to train models, especially when dealing with large datasets, is a significant obstacle. The rapid iterative creation of models is constrained by training times, which can grow significantly when dealing with a large dataset. Long training times further hinder the model's instant applicability when switching to real-time applications, making it less suited for time-sensitive scenarios like rapid diagnosis in healthcare settings. Another significant obstacle is the complex interplay between model complexity and memory allocation. The maximum batch sizes that can be utilized during training are constrained by memory capacity, which may reduce the model's ability to recognize complicated patterns in the data. These memory constraints could be made worse by complex models with many layers and parameters, which could exceed the amount of memory that is available. This can reduce the model's ability to capture subtle details, which will affect the accuracy of its classification and its ability to discriminate.

### A. Dataset-derived Limitations

The dataset's intrinsic biases have the potential to have a major impact on the effectiveness and generalizability of the model. These biases may result from a variety of causes, such as demographic, geographic, or historical ones. Biases may develop as a result of differences in patient demographics, scanning technologies, or data collection techniques across various geographic or temporal contexts. Such biases can have two negative effects: they can result in an unfair depiction of some classes, which distorts the model's impression of their predominance, and they can create an unnatural bias in favor of dominating classes.

A special set of difficulties are presented by the authenticity of medical images, which is essential for reliable diagnosis. Resolution, contrast, and noise variations in an image's quality can lead to complexities that prevent accurate classification. The model may lead to misdiagnosis due to poor image quality. As a result, misclassification is more likely to occur and the system is less resilient in real-world situations where image quality may be reduced.

Class distribution imbalances can have a major effect on how well a model performs after being trained. Situations in which particular classes are overrepresented constitute a serious problem. The model may develop a preference for overrepresented classes as a result of training on unbalanced datasets, potentially pushing minority classes to the edge of their discriminatory power. Additionally, the risk of misclassification in crucial medical contexts may increase as rare classes appear because of their dominating counterparts.

### B. Model Generalizability

The imperative evaluation of the model's behavior under scenarios of data variability beyond its training ambit assumes paramount significance. The model's performance may be disrupted when exposed to data distributions, imaging techniques, or clinical settings that differ significantly from its training environment. Its ability to identify strong features and patterns that go beyond the limitations of mixed datasets is what gives it the ability to extrapolate knowledge across diverse fields.

The flexibility of the model to adapt to unknown instances of unknown data is a need for the model to perform well. While some models exhibit remarkable transferability, others could require recurring retraining to consider the changing data landscape. The creation of systems for constant monitoring, recalibration, and harmonic integration within the changing therapeutic paradigm is also necessitated by the specter of obsolescence.

The empirical testbed of the model, set amongst the complicated reality of real-world circumstances, serves as the model's crucible for validation. The gap between theoretical power and practical application is highlighted by the disjunction in performance indicators when the model moves from the organized confines of controlled experiments to the nuanced details of real clinical usage. These inequalities highlight the importance of real-world evaluations, which provide essential information that informs model modifications and aligns the model's predictive power with the complex and ever-evolving landscape of clinical treatment.

### C. Deployment Issues

The model's capacity to adapt to a growing user base and expanding data streams will determine how well it is implemented. As demand grows, the difficulties lay in overcoming possible constraints linked to memory allocation, processing resources, and network bandwidth. To guarantee the model is capable of scaling, it is crucial to implement strong parallelization algorithms, load balancing systems, and use distributed computing paradigms. Additionally, effectively managing resources is required which may be addressed by designing an efficient architecture using microservices or containerization.

The model's performance is assessed in the setting of crucial medical decision-making based on its capacity to offer real-time insights, combining diagnostic precision with prompt response. To achieve real-time processing, careful optimization work must be done to reduce latency and processing delays. This necessitates thoughtful model complexity management, deliberate algorithmic optimization, and efficient feature extraction. It is crucial to strike a balance between model sophistication and computing efficiency to allow real-time functionality without sacrificing diagnostic efficacy.

Complex issues with interoperability and data management arise when the model is integrated into elaborate medical systems and databases. Version control processes, strong data pipelines, and rigorous data mapping are required to address compatibility difficulties and ensure data synchronization. Due to differences in data formats and technologies, compatibility problems could occur, making middleware development necessary to ensure smooth communication. To provide a dependable exchange, efficient data transport requires strong compression methods and low-latency protocols. Bidirectional data flow and versioning solutions are required to maintain coherence between the model and current systems due to synchronization complexity. Technical expertise and rigorous testing are vital for a successful deployment and for clinical decision-making.

A thorough evaluation of the model's performance has revealed its advantages and drawn attention to any weaknesses that need to be considered when interpreting results or considering its use in the real world. The difficulties mentioned—which include resource restrictions, limitations arising from datasets, and model generalizability—underline the necessity of taking a cautious and knowledgeable approach. Overcoming the stated constraints is of utmost significance for future studies. Resource limitations can be eased by investigating sophisticated parallel processing architectures and optimized algorithmic implementations. Model generalizability could be improved by using transfer learning approaches specifically designed for medical imaging data. For addressing class imbalances and improving overall performance, techniques like data synthesis and augmentation are crucial. Including clear AI approaches could improve the clinical interpretability of data. Research frontiers must be advanced by collaborative projects integrating machine learning experts, medical professionals, and subject specialists. By integrating these efforts, the field may overcome obstacles, cultivate reliable models ready for real-world use in healthcare settings, and ultimately lead to enhanced diagnostic accuracy and patient outcomes.

## XV. Conclusion

This investigation into breast cancer classification culminates in a model underpinned by machine learning methodologies, demonstrating a profound synergy between oncological diagnostics and computational techniques.

### A. Research Contributions and Discoveries

This study aimed to design an efficient deep-learning model for the classification of breast cancer images. By employing convolutional neural networks (CNNs) and optimizing various hyperparameters, a discriminative model was successfully trained to distinguish between normal, benign, and malignant tumor presentations. Key insights into the distribution of image intensities and their relation to tumor classifications were discerned. The deployment of the model into a Streamlit web application exemplifies how academic research can be translated into tangible, user-centric tools.

### B. Implications for Healthcare

The implications of this model are not limited to the domain of machine learning; they resonate profoundly within healthcare. With this enhanced diagnostic aid, clinicians can achieve quicker, more accurate diagnoses, expediting patient treatment plans. By automating a segment of the diagnostic process, the model facilitates optimal resource allocation in healthcare settings, fostering both efficiency and precision.

### C. Future Directions

The model's current performance, while commendable, provides a canvas for several future trajectories:

- *Model Optimization:* Continuous feedback from the deployed tool can help in gradient descent-based refinements, iteratively reducing loss and improving the model's precision and recall metrics.
- *Diverse Architectures*: While the current model leverages a particular CNN architecture, exploring alternative configurations, such as Residual Networks (ResNets) or Transformer-based models, could further elevate its performance.
- *Expanding Datasets:* One of deep learning's core tenets is its capacity to thrive on vast datasets. Incorporating larger and more diverse datasets can enhance the model's generalization, reducing overfitting and bolstering its robustness in varied diagnostic scenarios.
- *Adding further Classes:* It is evident that the dataset currently used is limited in the number of classifications it provides. A big step in this research would be to gain data that provides deeper classifications for example delving deeper into types of malignancies (adenocarcinoma, melanoma, etcetera) and types of benign tumors (lipomas, adenomas, etcetera). Another important aspect to note is datasets that contain unclassified images, or images that would be hard for even humans to identify. Having these broader class definitions will increase the robustness of the deployment significantly.

In summary, though it may be seen that this endeavor did not maintain levels of accuracy and other metrics such as this endeavor underscores the promise and potential of deep learning in oncological diagnostics. As the confluence of machine learning and healthcare continues to expand, there's great anticipation for even more refined tools and methodologies that can further the mission of early and accurate cancer detection.

ACKNOWLEDGMENT *(Heading 5)*

We would like to express our profound gratitude to both Mr. Pawel Pratyush and Dr. Sushant Kafle for their invaluable contributions to this project. Mr. Pratyush's guidance and mentorship were fundamental in bringing this project to fruition, and Dr. Kafle's expert guidance and meticulous editorial efforts greatly enhanced the quality of our paper. Their combined support and expertise were instrumental in the successful completion of this work, reflecting a collaborative effort we deeply appreciate.

REFERENCES

[1] World Health Organization. "Breast cancer." WHO, 2021. https://www.who.int/news-room/fact-sheets/detail/breast-cancer.

[2] Al-Dhabyani W, Gomaa M, Khaled H, Fahmy A. Dataset of breast ultrasound images. Data in Brief. 2020 Feb;28:104863. DOI: 10.1016/j.dib.2019.104863.

[3] Stark, G. F., Hart, G. R., Nartowt, B. J., & Deng, J. (2019). Predicting breast cancer risk using personal health data and machine learning models. PLOS ONE, 14(12). doi:10.1371/journal.pone.0226765.

[4] E. A. Bayrak, P. Kırcı and T. Ensari, "Comparison of Machine Learning Methods for Breast Cancer Diagnosis," 2019 Scientific Meeting on Electrical-Electronics & Biomedical Engineering and Computer Science (EBBT), Istanbul, Turkey, 2019, pp. 1-3, doi: 10.1109/EBBT.2019.8741990.

[5] LG, Ahmad, and Eshlaghy AT. "Using Three Machine Learning Techniques for Predicting Breast Cancer Recurrence." Journal of Health & Medical Informatics, vol. 04, no. 02, 2013, https://doi.org/10.4172/2157-7420.1000124.

[6] Gulzar, Yonis. "Fruit Image Classification Model Based on MobileNetV2 with Deep Transfer Learning Technique." Sustainability (Basel, Switzerland), vol. 15, no. 3, 2023, p. 1906–, https://doi.org/10.3390/su15031906.

[7] Hridayami, Praba, et al. "Fish Species Recognition Using VGG16 Deep Convolutional Neural Network." Journal of Computing Science and Engineering, vol. 13, no. 3, 30 Sept. 2019, pp. 124–130, https://doi.org/10.5626/jcse.2019.13.3.124. Accessed 20 Dec. 2020.

[8] Mi, Weiming, et al. "Deep Learning-Based Multi-Class Classification of Breast Digital Pathology Images." Cancer Management and Research, vol. 13, 2021, pp. 4605–17, https://doi.org/10.2147/CMAR.S312608.

[9] Vakili, Meysam, et al. "Performance Analysis and Comparison of Machine and Deep Learning Algorithms for IoT Data Classification." University of Science and Culture, 31 Jan. 2020. https://arxiv.org/ftp/arxiv/papers/2001/2001.09636.pdf

[10] Chicco, Davide, and Giuseppe Jurman. "The Advantages of the Matthews Correlation Coefficient (MCC) over F1 Score and Accuracy in Binary Classification Evaluation." BMC Genomics, vol. 21, no. 1, 2020, pp. 6–6, https://doi.org/10.1186/s12864-019-6413-7.

[11] Saleh, Hager, et al. "Predicting Breast Cancer Based on Optimized Deep Learning Approach." Computational Intelligence and Neuroscience, vol. 2022, 2022, pp. 1820777–11, https://doi.org/10.1155/2022/1820777.

[12] Mishra, A. K., Roy, P., Bandyopadhyay, S., & Das, S. K. (2021). Breast ultrasound tumour classification: A Machine Learning—Radiomics based approach. Expert Systems, 38(7), e12713. https://doi-org.libproxy2.usc.edu/10.1111/exsy.12713

[13] Cruz-Ramos C, García-Avila O, Almaraz-Damian JA, Ponomaryov V, Reyes-Reyes R, Sadovnychiy S. Benign and Malignant Breast Tumor Classification in Ultrasound and Mammography Images via Fusion of Deep Learning and Handcraft Features. Entropy (Basel). 2023 Jun 28;25(7):991. doi: 10.3390/e25070991. PMID: 37509938; PMCID: PMC10378567.

[14] Labcharoenwongs P, Vonganansup S, Chunhapran O, Noolek D, Yampaka T. An Automatic Breast Tumor Detection and Classification including Automatic Tumor Volume Estimation Using Deep Learning Technique. Asian Pac J Cancer Prev. 2023 Mar 1;24(3):1081-1088. doi: 10.31557/APJCP.2023.24.3.1081. PMID: 36974564; PMCID: PMC10334094